# Dielectric Modeling of Oil-paper Insulation Systems at High DC Voltage Stress Using a Charge-carrier-based Approach

**Tobias Gabler, Karsten Backhaus** and **Steffen Großmann**

Technische Universität Dresden, Institute of Electrical Power Systems and High Voltage Engineering

Mommsenstrasse 10

01062 Dresden, Germany

**Ronny Fritsche**

SIEMENS AG, Transformers

Katzwanger Strasse 150

90461 Nuremberg, Germany

## ABSTRACT

**It is state-of-the-art to describe the dielectric behavior of an insulation material by its permittivity and its specific electric conductivity in order to estimate the dielectric stress of an insulation system. Thus, the electric field at DC voltage stress is determined according to the stationary electrical conduction field with the electric conductivity of the insulation materials. However, at oil-insulated arrangements a higher field strength in front of bare metal electrodes at high DC voltage stress occurs, which is not explainable with this model. Therefore, a charge carrier-based approach is presented to describe the dielectric behavior of the oil-paper insulation. It describes the movement of charge carriers and their effect on the electric field strength. Their drift leads to an accumulation of charge carriers in front of electrodes which results in a higher field strength in these areas, which can be calculated numerically using the Poisson-Nernst-Planck equation system. Compared to the conductivity-based model fundamental differences can be shown. Breakdown experiments qualitatively confirm the expectations according to the charge carrier-based approach. The results show that the charge carrier-based field distribution has to be considered for modelling the electrical field strength distribution at high DC voltage stress. They also show, that the dielectric behavior of these arrangements cannot be explained according to the state-of-the-art model.**

## 1 INTRODUCTION

**HVDC** transmission is gaining in importance due to the increasing demand on transmitting a high amount of electrical energy over long distances. Because of the lower electric losses, it is economically more efficient than AC transmission at high distances [1]. Converter transformers play a key role by forming the interface between the HVAC and HVDC transmission e. g. by adapting voltage from the connected AC grid and by separating the rectifiers galvanically from the three-phase supply. During operation a superimposed DC voltage stress occurs on the secondary circuit of these transformers [2]. Hence, the understanding of the dielectric behavior of oil-paper insulation systems at DC voltage stress needs to be enhanced in order to improve the dielectric design of HVDC transformers.

It is state of the art to model an insulation system by an electrical network of *RC* components (Figure 1). This model considers the different polarization mechanisms ($R_{P1}$ and $C_{P1}$; … $R_{Pi}$ and $C_{Pi}$) of each material, its geometrical capacity $C_0$ caused by its permittivity $\varepsilon$ as well as its specific electric conductivity $\kappa$ ($R_\infty$). Thus, at DC voltage stress the distribution of the electrical field strength is determined according to the stationary electrical conduction field using the specific electric conductivity.

However, these assumptions are not sufficient for modeling the effects in gaseous, fluid or solid dielectrics at high DC voltage stress [3]. Furthermore, investigations showed an influence of the oil volume in the test vessel on the breakdown voltage of an opened arrangement [4], an inhomogeneous electrical field strength distribution and a higher field strength in front of electrodes at a homogenous electrode arrangement [5–7] and a non-linear conductivity of mineral oil at high DC voltage stress [8, 9]. Therefore, a charge carrier-based approach was presented [2, 10, 11].

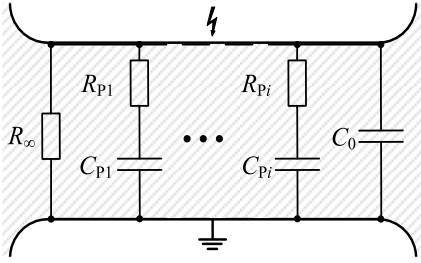

**Figure 1.** Equivalent circuit of a dielectric for modeling its dielectric behavior (*RC*-Model).

# 2 CHARGE CARRIER-BASED APPROACH

## 2.1 PHYSICAL FUNDAMENTALS

According to the charge carrier-based approach there are four species of charge carriers: oil intrinsic and injected charge carriers of each polarity [2, 11], which move through the oil at DC voltage stress due to electrostatic forces and diffusion processes. Oil intrinsic charge carriers dissociate from organic salts or acids dependent on the electrical field strength [11–13]. They are stable due to their micelle structure and hardly recombine [2, 11], which is why they can accumulate at electrodes (*hetero charges*). Injected charge carriers (*homo charges*) such as electrons and holes are provided by electrodes by electron attachment or detachment dependent on the local field strength in front of the electrodes [2, 11, 14]. In contrast to *hetero charges* they recombine with the oppositely charged electrode or mutually in oil with oppositely charged *homo charges*. In paper barriers, the mobility of charge carriers is reduced due to the higher density of paper compared to oil [2]. This can lead to an accumulation of charge carriers in front of paper barriers [2, 15, 16] or within the paper. Thus, the charge carrier-based, non-linear conductivity can be described by three main mechanisms (Table 1).

**Table 1.** Mechanisms of charge carrier-based conduction processes according to [2, 11].

| *Generation* | *Transport* | *Recombination* |
|---|---|---|
| Oil intrinsic charge carriers | Movement due to electrostatic forces | At bare electrodes |
| Charge carriers injected from electrodes | Diffusion processes | Mutually, dependent on particle density |

The accumulation of *hetero charges* at the oppositely charged electrodes locally increases the electrical field strength. As a result, their accumulation affects the injection of *homo charges*. Moreover, injected *homo charges* decrease the field strength at the injecting electrodes and thus influence the electric field strength as well. Hence, each charge carrier species has to be considered in a numerical model. [2]

The movement of these charge carriers through the dielectric causes a measurable current. It is assumed, that this current mainly consists of two different shares caused by the movement of *hetero charges* and the injection and movement of *homo charges* (Figure 2) [2]. After the voltage rising at $t_0$ and its caused displacement current at the arrangement, the current mainly decreases due to the movement of *hetero charges* (intrinsic current) to the oppositely charged electrode. In addition, there is an increasing current share due to the injection of *homo charges* (injected current). The accumulation of *hetero charges* at the electrodes leads to an increasing local field strength and thus to an increasing injected current. Furthermore, the increasing injection of *homo charges* slightly decreases the local field strength at the injecting electrode [2], which is why the increase of the injected current slightly decreases. After accumulation of *hetero charges* is finished ($t_1$), injection and recombination of *homo charges* and their influence on the local field strength reaches its equilibrium ($t_{stat}$). Thus, a stationary current indicates the equilibrium in the injection and recombination and the steady state of movement of *homo charges*, and thus the steady state in the electrical field strength distribution [2].

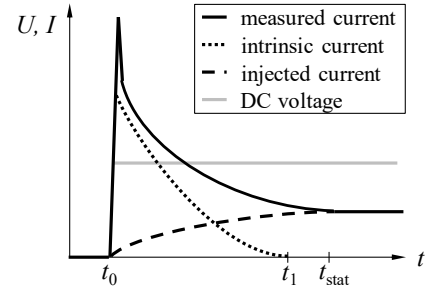

**Figure 2.** Schematic composition of polarization current caused by different species of charge carriers.

## 2.2 POISSON-NERNST-PLANCK EQUATION SYSTEM

The Poisson-Nernst-Planck equation is state of the art in modelling the behavior of the non-linear conductivity of semiconductors with lower charge carrier densities than ohmic materials. It is a coupled, non-linear, second order differential equation system. It consists of the Poisson Equation (1), which describes the electric field due to the potential $\varphi$ considering the permittivity $\varepsilon_0 \cdot \varepsilon_r$ of the material and the charge density, which is described as the product of the elementary charge e and the sum of the concentration of each charge carrier species (product of concentration $n$ and valence number $z$ of each species) [2].

The Nernst-Planck equations describe the local temporal changes of each charge carrier species $i$ by the equilibrium of the particle motion, Equation (2).

$$-\Delta \varphi = \frac{e}{\varepsilon_0 \varepsilon_r} \sum z_i n_i \qquad (1)$$

$$\frac{\partial n_i}{\partial t} = z_i \mu_i n_i \Delta \varphi + D_i \Delta n_i + \sum (r_{i\ ion} + r_{ij\ rec}) \qquad (2)$$

where:
- $[n_i]$ … Concentration of charge carriers of species $i$
- $[z_i]$ … Valence number of species $i$
- $[\mu_i]$ … Mobility of charge carriers of species $i$
- $[D_i]$ … Diffusion coefficient of species $i$
- $[r_{i\ ion}]$ … Ionization rate of species $i$
- $[r_{ij\ rec}]$ … Recombination rate of species $i$ with species $j$

The local density is influenced by electrostatic forces (1st addend) and diffusion processes (2nd addend). The source terms $r_i$ (3rd addend) describe the time dependent change of ion concentration due to ionization or recombination processes [2]. The product of $n_i \cdot \Delta\varphi$ is called a strong coupling [17], which demands a numerical calculation.

## 2.3 BOUNDARY CONDITIONS

In order to calculate the distribution of electrical field strength, several boundary conditions have to be set based on the physical assumptions [2, 11].

The electrical potential of the electrodes has to be set according to the test voltage, which must be time dependent, thus

$$\varphi_{HV}(t) = u_{test}(t) \quad (3)$$

$$\varphi_{GND}(t) = 0 \quad (4)$$

As a simplification, the sum of *hetero charges* is assumed to be constant. Thus, the generation or recombination of *hetero charges* in oil or at the electrode surface is neglected (Equations (5) and (6)), knowing that there has to be a generation [12] and a slight recombination of *hetero charges*.

$$n_i(t=0) = n_{0i} \quad (5)$$

$$z_i \mu_i n_i \Delta\varphi - D_i \Delta n_i = 0 \quad (6)$$

*Homo charges* are generated at the corresponding electrode [14]. Investigations showed an analogy to the Fowler-Nordheim electron emission from metal into vacuum [9, 18], which is strongly dependent on the local field strength $E$ at the electrodes (Equation (7)). In any case, the injection must be monotonously increasing dependent on the field strength,

$$z_i \mu_i n_i \Delta\varphi - D_i \Delta n_i = a_i \cdot E^{b_i} \cdot \exp(^{-c_i}/_E) \quad (7)$$

Due to the missing micelle structure and the excessive charge in terms of the octet rule of the injected charge carriers, *homo charges* recombine at the opposite electrode (equation (8)) and mutually in the oil with the opposite species, dependent on their local concentration.

$$z_i \mu_i n_i \Delta\varphi - D_i \Delta n_i = -\mu_i n_i E \quad (8)$$

# 3 COMPARISON OF FIELD CALCULATIONS ACCORDING TO CHARGE CARRIER-BASED APPROACH AND *RC* MODEL

In order to confirm the charge carrier-based approach simple oil and oil-paper insulations are used. To reduce calculation time and to compare the calculated results with experiments one-dimensional arrangements were built which correspond to a homogenous electrode arrangement at high DC voltage stress as used in the breakdown experiments (cf. Figure 7). The distribution of the electrical field strength of these arrangements in stationary conditions will be compared qualitatively in the following according to the numerical calculations. Comparing the conductivity-based model ($\kappa$) or the charge carrier-based approach (PNP) fundamental differences in the distributions of the electrical field strength can be shown.

## 3.1 BARE ELECTRODE ARRANGEMENT

At the bare electrode arrangement, an accumulation of intrinsic charge carriers will take place at the electrodes (Figure 3). These charge carriers with opposite charge – *hetero charges* – will lead to an increased field strength in front of the electrodes. In addition, the high field strength in front of the electrodes influences the injection of charge carriers [19], which will decrease the electrical field strength [2]. Thus, the local maximum of the field strength in front of the electrodes is highly dependent on the amount of accumulated *hetero charges*, i. e. a higher amount of *hetero charges* leads to a higher field strength in front of the electrodes. However, the geometrical field distribution would assume a homogenous field according to the state-of-the-art modeling. [15, 16]

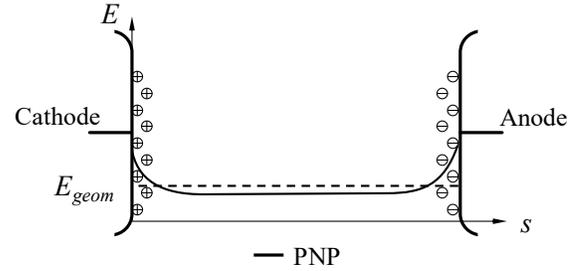

**Figure 3.** Qualitative comparison of field strength distribution of bare arrangement due to the geometry of the arrangement ($E_{geom}$) and the PNP approach according to [16].

## 3.2 PAPER INSULATION AT BOTH ELECTRODES

According to the charge carrier-based approach, the main stress is expected in the oil gap, if both electrodes of the arrangement are paper-insulated. Due to the decreased mobility of charge carriers in the paper [2], there is a high concentration of *homo charges* in the paper barriers with the same polarity as each electrode (Figure 4). This result in an almost homogenous field distribution in the oil gap and, in addition, in a lower maximum field strength compared to the bare arrangements. Hence, higher breakdown voltages should be obtained during the breakdown experiments and in addition, a dependency of the breakdown voltage on the oil gap has to take place. [15, 16] Furthermore, the polarity effect in the charge carrier injection [9] leads to a different field distribution in the paper insulation on front of the cathode and the anode.

In contrast to the charge carrier-based approach, the main stress would occur in the paper barrier according to the conductivity-based model. Therefore, the breakdown voltage would be only slightly dependent from the oil gap, which is why the breakdown would initiate in the paper [15, 16].

## 3.3 BARE ELECTRODES AND THREE OIL GAPS

At the arrangement with bare electrodes, two centered paper barriers and three oil gaps, the expected stress of the centered oil gap between barriers B1 and B2 is similar compared to the arrangement with both electrodes paper-insulated (Figure 5).

As well as at the bare arrangement, *hetero charges* would accumulate at the bare electrodes, which locally leads to a slightly higher field strength in front of the electrodes and thus to an increased injection of *homo charges*. Hence, these *homo*

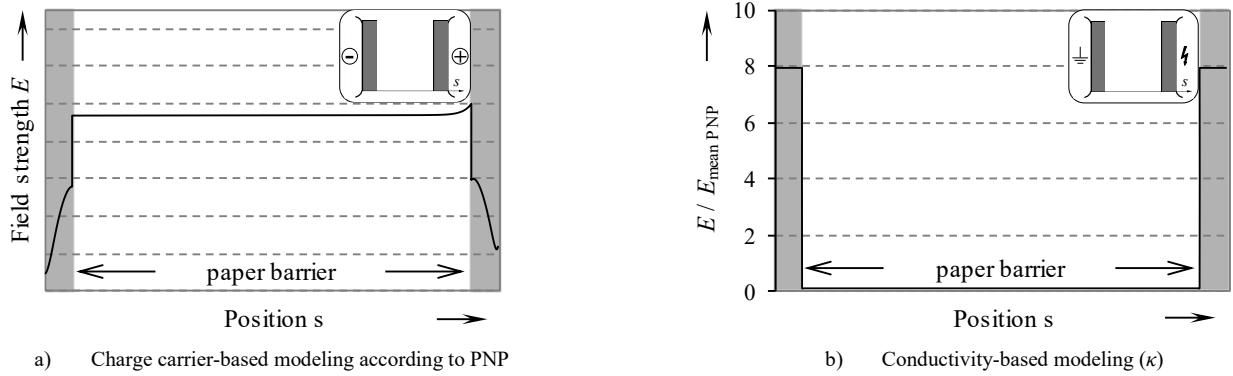

a) Charge carrier-based modeling according to PNP

b) Conductivity-based modeling ($\kappa$)

**Figure 4.** Calculated field strength distribution of arrangement with insulated electrodes, comparison between PNP (a) and $\kappa$ (b).

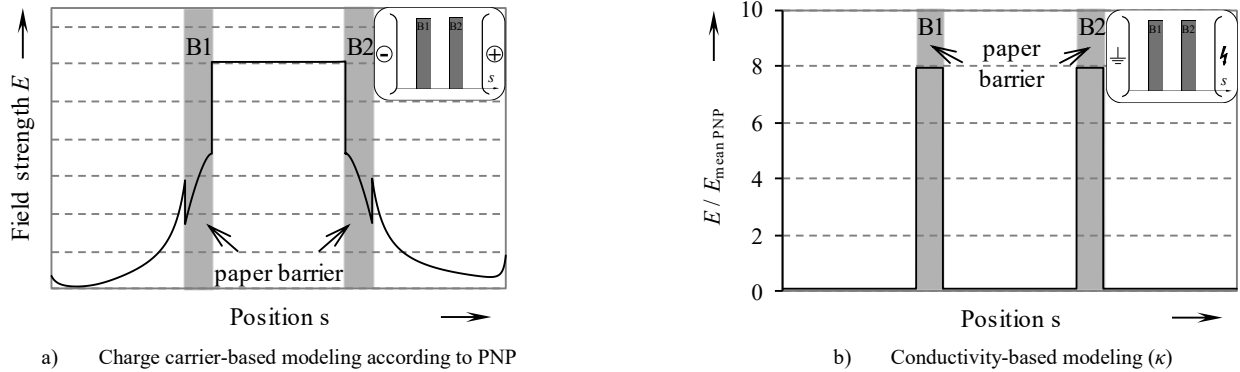

a) Charge carrier-based modeling according to PNP

b) Conductivity-based modeling ($\kappa$)

**Figure 5.** Calculated field strength distribution of arrangement with three oil gaps, comparison between PNP (a) and $\kappa$ (b).

*charges* will mainly concentrate in the paper barriers, which decreases the dielectric stress of the oil gaps in front of each electrode. In consequence, the main dielectric stress occurs in the centered oil gap. Due to the much lower field strength in the outer gaps, the breakdown strength of the arrangement should highly depend on the width of the centered oil gap and has to be almost independent of the outer oil gaps. The polarity effect in the charge carrier injection [9] leads to a different field distribution on front of anode and cathode. Investigations at other arrangements with multiple oil gaps show similar results with higher field strengths in the centered oil gaps as well [20].

In contrast, almost no dependency on the oil gap would occur according to the conductivity-based model as well as shown at the arrangement with paper-insulated electrodes.

## 4 EXPERIMENTAL SETUP

### 4.1 HIGH VOLTAGE TEST CIRCUIT

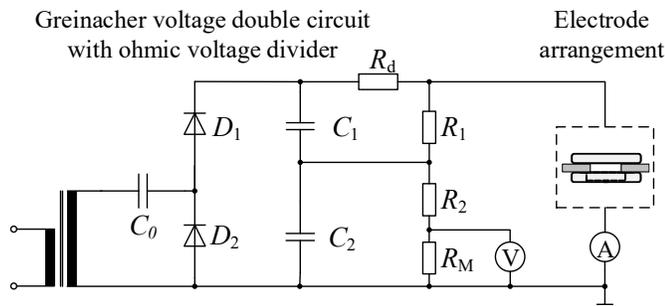

**Figure 6.** High voltage test circuit.

A Greinacher voltage double circuit supplies the test arrangement with high DC voltage up to $V_{DC}$ = 250 kV.

The voltage is measured using an ohmic voltage divider. Furthermore, a damping resistance $R_D$ connects the ohmic voltage divider and the voltage double circuit (Figure 6) to avoid root points on the electrodes after breakdown, which may influence following experiments.

### 4.2 ELECTRODE ARRANGEMENT

The breakdown experiments were performed at a circular plane-parallel electrode arrangement made of high-grade steel. The arrangement is located on a grounded base slab in a test vessel filled with mineral oil (Figure 7).

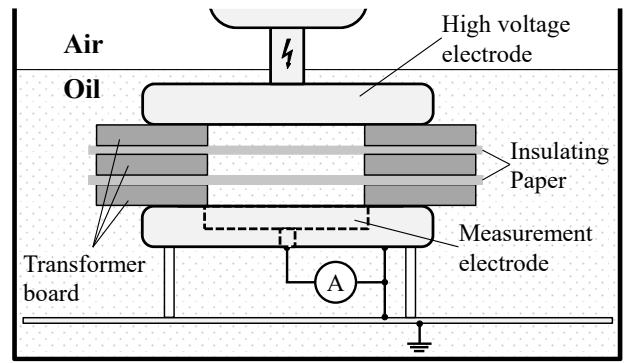

**Figure 7.** Schematic diagram of test arrangement in test vessel (exemplary at arrangement with three oil gaps) according to [16].

In the investigations, the experiments with insulated electrodes or paper barriers are performed with high-density insulation paper of 0.13 mm thickness. Investigations showed a dependency of the breakdown voltage of an opened arrangement on the oil level in the test vessel [4] (cf. Sections 5 and 6.1). Therefore, the arrangement is sealed with barriers made of transformerboard in order to avoid an influence of the surrounding oil on the breakdown experiments (Figure 7).

Simple oil- and oil-paper-insulated arrangements according to section 3 were examined to confirm the charge carrier-based approach. The breakdown experiments were performed using arrangements with bare electrodes and paper-insulated electrodes. At the bare arrangement, the influence of the surrounding oil is exemplified. The oil gap width was set between (3 … 10) mm by the thickness of the used transformer board. The paper barriers and the sealing boards are fixed by the weight of the high voltage electrode.

Additionally, experiments at arrangements with bare electrodes, two paper barriers and three oil gaps (Figure 5 and Figure 7) are performed to confirm the expectations according to Section 3.3. The oil gaps of this arrangement are set to achieve same gap widths in the center and in front of the electrodes (A, B) or same gap widths in the center with different oil gap widths in front of the electrodes (B, C; Table 2) [16].

**Table 2.** Arrangements with three oil gaps and two paper barriers according to Figure 5.

| Setup | $s_{Cath-B1}$ | $s_{B1-B2}$ | $s_{B2-An}$ |
|---|---|---|---|
| A | 1 mm | 1 mm | 1 mm |
| B | 3 mm | 3 mm | 3 mm |
| C | 1 mm | 3 mm | 1 mm |

### 4.3 CURRENT MEASUREMENT SETUP

The movement of intrinsic and injected charge carriers in the arrangement as well as their generation result in a polarization current which can be measured during the experiments (cf. Section 2.1). Hence, the steady state of movement and diffusion processes can be determined by a current measurement to compare the experimental results with field calculations. To perform current measurements during the experiments a guard ring arrangement is used as ground electrode (Figure 7).

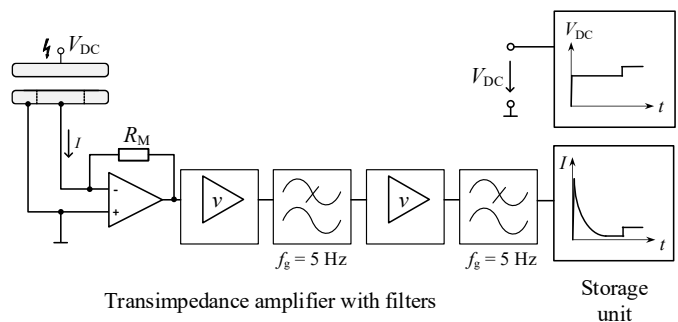

**Figure 8.** Current measurement setup.

A transimpedance amplifier measures this current, which is connected directly to the guard ring arrangement (Figure 8). Different measuring ranges can be set by the resistor $R_M$ in the feedback loop of the amplifier. Low-pass filters and amplifiers are used to avoid disturbing influences due to the ripple of the high voltage test circuit on the expected direct current. Via Bluetooth connection the measured voltage and current at the arrangement can be visualized and stored using a PC simultaneously during the breakdown measurements.

### 4.4 EXPERIMENTAL PROCEDURE

In order to provide consistent conditions at each breakdown experiment, the used oil, the paper barriers and the transformer board sealing need to be prepared properly [2, 15]. To ensure a consistent quality with a maximum moisture content of 7 ppm the used mineral oil is vacuum dried and degassed at a maximum pressure of $p_{abs}$ = 100 Pa using an oil purifying plant. The used paper barriers as well as the transformer board are prepared according to IEC 60763-2 and are vacuum dried in a heating cabinet for at least 24 h at 105 °C at a pressure of $p_{abs}$ = 100 Pa. After the drying process, the dried samples are impregnated with the insulating oil and kept under vacuum in a vacuum chamber for storage.

Breakdown experiments are performed with a stepped voltage rising test with simultaneous current measurements. In order to confirm the expectations according to the charge carrier-based approach it is necessary to ensure that polarization and drift processes have finished before increasing the voltage. Therefore, the voltage will not be increased until a stationary current is achieved. The starting voltage of the experiments is set at approx. 75 % of the expected breakdown voltage. Depending on the examined arrangement, gap distance and starting voltage the duration of each voltage step is about (10 … 60) min (cf. section 5). Due to symmetry of the arrangements breakdown experiments are performed at positive DC voltage only. [2, 9, 16]

## 5 CURRENT MEASUREMENT RESULTS

The results of the current measurement during the breakdown experiments show a strong difference between the opened and the sealed arrangement (Figure 9).

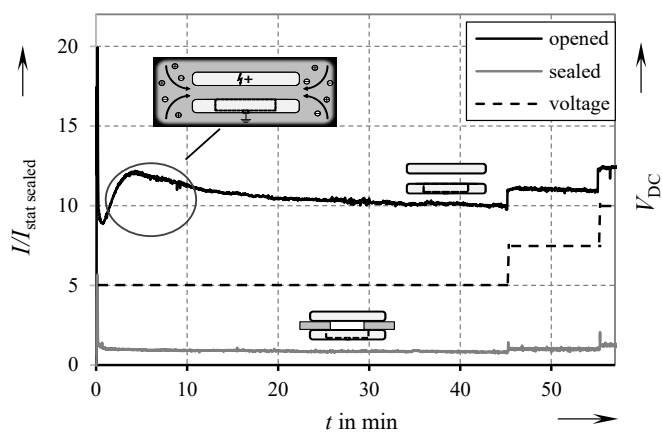

**Figure 9.** Exemplary measurement of time-dependent current $I$ of the opened and sealed bare arrangement.

After the first displacement current caused by the initial voltage rise, a decreasing polarization current can be measured in the first few minutes of the measurements at the opened arrangement. After this first decreasing current, another increase is measured over a few minutes, which subsequently decreases after reaching its maximum. This second decrease is explainable with additional "parasitic" charge carriers. At the opened arrangement the electrodes work as an electrostatic filter and thus additional *hetero charges* from the surrounding oil volume get into the arrangement. Consequently, an increasing current can be measured. Furthermore, these additional charge carriers accumulate at the electrodes, increase the field strength and thus increase the injection of *homo charges* of the arrangement. In addition, also *homo charges*, injected from "parasitic" electrodes, e.g. bare base slab or outer electrode compartments, move to the measurement electrode and can be measured as an increasing current [9]. Due to these additional charge carriers from the surrounding oil volume, reaching the equilibrium of accumulated intrinsic charge carriers and the steady state of the injection of charge carriers take a long time. Thus the current takes up to (30 … 60) min to reach a steady state because of their mutual influence on the local field strength (cf. Section 2.1).

At the sealed arrangement, the polarization current also reaches its steady state after a few minutes. But there is no second increase measurable. Due to the inserted mechanical barrier around the arrangement, almost no additional charge carriers from the surrounding oil volume can move inwards. These results show that charge carriers from the surrounding oil do not affect the sealed arrangement. Moreover, the stationary current at this arrangement is much lower than the current of the opened arrangement in stationary conditions. Only *hetero charges* from the oil volume inside the barrier accumulate at the electrodes, which leads to lower field strength at the electrodes and thus to a lower current caused by *homo charges* (cf. Section 2.1). In addition, no *homo charges* from the surrounding can affect the measured current.

Furthermore, after the first stationary current no further polarization current is measured after the next voltage step at both arrangements. It shows that increasing the voltage only increases the field strength within the arrangement and thus the current caused by *homo charges* increases. However, no *hetero charges* are involved in this current because all *hetero charges* accumulated during the first voltage step. Thus, no decreasing current share occurs. In contrast to the modeling according to the state-of-the-art model, further polarization currents should occur after each voltage rise.

Another influence of the moving charge carriers can be shown by the movement of the oil within the test vessel during the experiments [9, 15]. Moreover, an intense movement of the oil at the high voltage connection was observed (Figure 10), which is much stronger at the opened arrangement compared to the sealed arrangement. The reason of this ascending oil at the bare connection is explainable by the movement of charge carriers, which are bounded to oil molecules [2]. *Hetero charges* accumulate at bare metallic parts during the experiment such as electrodes or the high voltage connection. This result in a movement along the surface due to electrostatic forces to other bare parts where no charge carriers have accumulated at this time [2]. Hence, oil can ascend along the bare high voltage connection at the oil-air interface.

The reason of the stronger movement at the opened arrangement is the higher amount of charge carriers in the whole oil volume compared to the sealed arrangement. Due to the high electrical field strength between the electrodes, more *hetero charges* can be formed and more *homo charges* are injected. They can leave the opened electrode arrangement and ascend at the high voltage connection very high [2, 15]. After a few minutes, the oil descends again, because accumulated charge carriers slightly recombine. This descend also correlates with the decreasing of the measured current through the opened arrangement. Therefore, it is assumed that the descending oil is a result of *hetero charges*, which may recombine with *homo charges*.

At the sealed arrangement the charge carriers of the electrode arrangement cannot leave the arrangement. Accordingly, only the 'parasitic' charge carriers from the surrounding oil volume can accumulate at the high voltage connection. In addition, there is a lower amount of charge carriers in the surrounding oil volume due to the much lower field strength at the 'parasitic' electrodes. Hence, a fewer amount of oil can ascend at the bare metallic high voltage connection. [15]

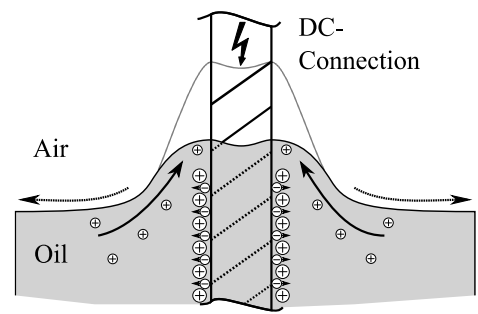

**Figure 10.** Ascending oil at high voltage connection at negative DC Voltage stress according to [2].

# 6 RESULTS OF BREAKDOWN EXPERIMENTS

## 6.1 OPENED AND SEALED BARE ARRANGMENT

The results of the breakdown experiments show a strong difference (Figure 11) between the opened and sealed, bare electrode arrangement (b-b) [15]. Thus, a strong influence of the surrounding oil volume on the breakdown strength can be measured in the experiments as shown at the current measurements as well. At the opened arrangement with bare electrodes, much lower breakdown voltages can be measured compared to the sealed arrangement.

These results show that the charge carriers from the arrangement and, in addition, the charge carriers from the surrounding oil volume strongly influence the field strength at the electrodes and thus the breakdown voltage of the arrangement [15]. At the opened arrangement, a much higher amount of *hetero charges* can accumulate at the electrodes. Consequently, the maximum electrical field strength in front of the electrodes is much higher than the field strength at the sealed arrangement, which leads to a lower breakdown strength of the opened arrangement. In addition, also the injected charge carriers of the 'parasitic' electrodes from the surrounding oil influence the field strength at the electrodes. Furthermore, the increase of the breakdown voltages with increasing oil gap witdth of the opened arrangement is similar to the increase of the breakdown voltage of the sealed, bare arrangement. It shows that no additional effects influence these results, e. g. the longitudinal dielectric interface of the sealed arrangement or a volume effect due to the reduced electrode area. According to the currently used conductivity-based modelling, almost no dependency of the surrounding oil should occur and similar breakdown voltages would be expected at the sealed and the opened arrangement.

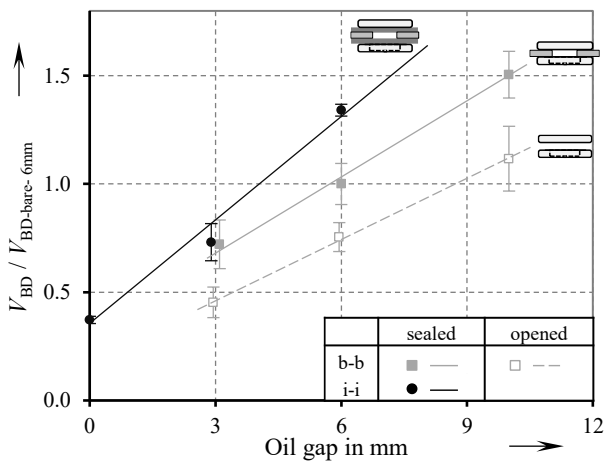

**Figure 11.** Results of breakdown measurements at opened and sealed oil- and oil-paper-insulated arrangements.

## 6.2 OIL AND OIL-PAPER-INSULATED ARRANGEMENTS

The comparison of the breakdown voltages of the sealed bare (b-b) and paper-insulated arrangements (i-i) confirm the expectations based on the charge carrier-based approach (Figure 11). At the arrangement with both electrodes paper-insulated, higher breakdown voltages are measured compared to the bare arrangement.

The paper insulation is filled with injected *homo charges* with the same polarity as the insulated electrode in the steady state due to the reduced mobility of the charge carriers in paper barriers [2] and thus injection of charge carriers from the insulated electrodes is reduced. As a result, the potential of the electrodes is shifted to the oil-paper interface. This leads to a lower field strength in the paper barrier. Hence, there is an almost homogenous distribution of the electrical field strength and a much lower maximum field strength in the oil gap of the paper-insulated arrangement compared to the bare arrangement (Figure 4). This leads to higher breakdown voltages as measured. Therefore, a strong dependency of the breakdown voltage on the oil gap width can be identified due to the main dielectric stress in oil. [16]

According to the conductivity-based model, only a slight dependency of the breakdown voltage on the oil gap width should occur because the main dielectric stress is in the paper insulation (Figure 4). Thus, the results confirm the charge carrier-based approach qualitatively and do not correspond to the conductivity-based model. [16]

## 6.3 ARRANGEMENT WITH THREE OIL GAPS AND TWO PAPER BARRIERS

The charge carrier-based approach can also be confirmed at the arrangement with bare electrodes and three oil gaps [16]. The results of the arrangements with increased oil gaps (A; B) show that increasing the oil gap width leads to an increased breakdown voltage (Figure 12). Hence, the main stress occurs in the oil gap instead of the paper insulation, which does not correspond to the conductivity-based model as well (Figure 5). Due to this model, almost the same field strength should occur in the paper barrier at these arrangements. Thus, no dependency of the breakdown voltages on the oil gap width should take place.

Furthermore, the outer oil gaps do not affect the breakdown voltage of the arrangement. The results of setups with an inner oil gap width of $s_{B1\,B2} = 3$ mm (B; C) are similar, while the width of the outer oil gap of the arrangement was decreased (Figure 12). Analogous to the arrangement with paper-insulated electrodes the paper barriers and the outer oil gaps between the bare electrodes and the paper barriers are filled with *homo charges* with the same polarity as the corresponding electrode. Hence, the potentials of the electrodes are 'shifted' to the inner barriers. This leads to a lower field strength in the outer oil gaps, which is why the main dielectric stress occurs in the inner gap $s_{B1\,B2}$ (Figure 5). Therefore, the dielectric strength of the arrangement is almost independent of the width of the outer oil gaps, which leads to similar results for the arrangements B and C. Furthermore, every root point in the paper barriers are oriented from the inner oil gap to the electrodes. It also shows that the breakdown starts in the inner oil gap as expected according to the charge carrier-based approach [16].

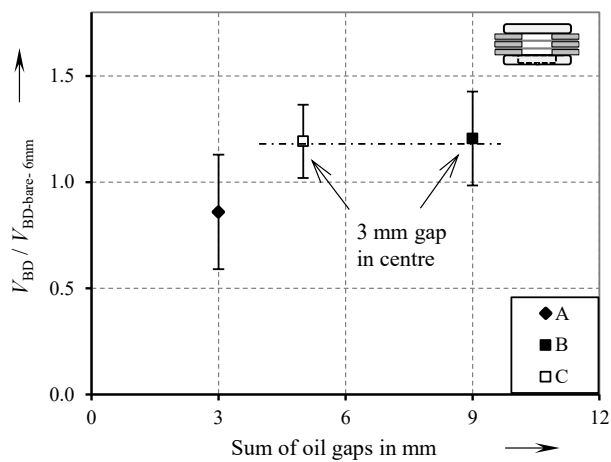

**Figure 12.** Breakdown results of arrangement with three oil gaps according to Table 2 [16].

## 7 SUMMARY

In this paper, the authors showed a validation of a charge carrier-based approach to describe the dielectric behavior of oil- and oil-paper-insulated arrangements at high DC voltage stress. This stress leads to a movement of intrinsic charge carriers and charge carriers, which are injected from electrodes or metallic parts respectively. Due to this movement, charge carriers can accumulate in front of electrodes or at paper barriers. These physical processes can be simulated by the Poisson-Nernst-Planck equation system, which shows strong differences in the distribution of the electrical field strength compared to the state-of-the-art $\kappa$-modelling. Breakdown experiments and current measurements showed a qualitatively good accordance to the results according to the charge carrier-based approach. Moreover, the experiments showed that the movement of every species of charge carriers has to be taken into account while doing experimental investigations at high DC voltage stress. Furthermore, the authors can show that modeling the dielectric behavior of oil-paper-insulated arrangements at high DC voltage stress according to the specific electric conductivity $\kappa$ leads to insufficient results.

## ACKNOWLEDGMENT

The authors gratefully thank SIEMENS AG, Division Transformers in Nuremberg for funding this research cooperation over the past years.

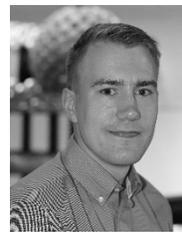
**Tobias Gabler** was born in Freiberg, Germany in 1989. He received the diploma degree in electrical engineering from the Technische Universität Dresden, Germany in 2015. Since 2015, he has been a research assistant and PhD student at the Institute of Electrical Power Systems and High Voltage Engineering at Technische Universität Dresden. His research interest is the electrical field strength distribution of oil-paper insulation systems at HVDC stresses. He is student member of CIGRÉ and a member of the German VDE. In 2017 he received the "Best Student Paper Award" of the ISH 2017.

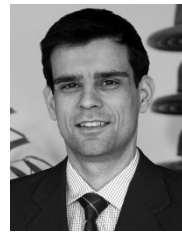
**Karsten Backhaus** was born in Riesa, Germany in 1981. He received the diploma thesis in mechanical engineering in 2007 and the Ph.D. degree in electrical engineering from the Technische Universität Dresden, Germany in 2016. Since 2008, he has been a research assistant at the Institute of Electrical Power Systems and High Voltage Engineering at Technische Univer-sität Dresden. His research interests include field grading materials for corona protection and oil-paper insulations. Karsten Backhaus received the "John Neal Award" of the EEIM in 2015 and the "Innovationspreis des Industrieclubs Sachsen e.V." in 2017.

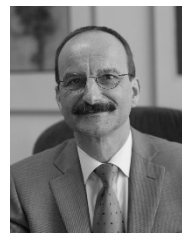
**Steffen Großmann** was born in Dresden, Germany in 1954. He received the diploma and Ph.D. degree in electrical engineering from the Technische Universität Dresden, Germany in 1976 and 1988. He started working for Richard Bergner GmbH in 1990, dealing with electrical and mechanical behavior of fittings for substations and overhead lines. In 1997 he became product team manager for substations and low voltage materials. Since 2004 he has been full professor in the field of electrical power systems and high voltage engineering at Technische Universität Dresden.

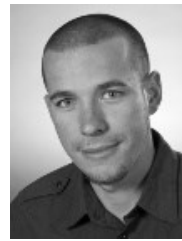
**Ronny Fritsche** was born in Löbau, Germany in 1981. He received the diploma degree in electrical engineering from the Zittau/Görlitz University of Applied Sciences, Germany in 2006. Since 2006 he has been working for SIEMENS AG. Since 2009 he has been working as an engineer for R&D of power transformers and in 2013 he became head of R&D Power Transformers in Nuremberg, Germany. He is CIGRÉ active member of working groups JWG A2/D1.41; WG D1.70; WG A2.59 and WG D1.63.